\documentclass[twocolumn]{autart}    

\usepackage{graphicx}          

\usepackage{amsmath}

\usepackage{amsfonts}
\usepackage{mathabx}
\usepackage{color}
\usepackage{epstopdf} 
\usepackage{tikz}
\usepackage{mathtools}

\newcommand{\smallsub}[1]{\! {\scriptscriptstyle \mathcal{#1}} }

\newtheorem{remark}{Remark}

\newtheorem{property}{Property}

\DeclarePairedDelimiter\ceil{\lceil}{\rceil}
\DeclarePairedDelimiter\floor{\lfloor}{\rfloor}

\newcommand{\Eb}{\bar{\mathbb{E}}}

\newcommand{\E}{\mathbb{E}}

\newcommand{\mD}{\mathcal{D}}

\newcommand{\mN}{\mathcal{N}}

\newcommand{\mR}{\mathcal{R}}
\newcommand{\smR}{\smallsub{R}}

\newcommand{\smD}{\smallsub{D}}

\newcommand{\beq}{\begin{equation}}
\newcommand{\eeq}{\end{equation}}
\newcommand{\beqs}{\begin{equation*}}
\newcommand{\eeqs}{\end{equation*}}
\newcommand{\beqr}{\begin{eqnarray}}
\newcommand{\eeqr}{\end{eqnarray}}
\newcommand{\beqrs}{\begin{eqnarray*}}
	\newcommand{\eeqrs}{\end{eqnarray*}}

\usepackage{algorithm}
\newcommand{\norm}[1]{\| #1 \|}

\usepackage{cite}

\begin{document}

\begin{frontmatter}

\title{A frequency domain approach for local module identification in dynamic networks \thanksref{footnoteinfo}} 

\thanks[footnoteinfo]{This project has received funding from the European Research Council (ERC), Advanced Research Grant SYSDYNET, under the European Union’s Horizon 2020 research and innovation programme (Grant Agreement No. 694504), and Strategic Research Program SRP60 of the Vrije Universiteit Brussel.}

\author[Tue]{Karthik R. Ramaswamy}\ead{k.r.ramaswamy@tue.nl},    
\author[Vub]{P\'{e}ter Zolt\'{a}n Csurcsia}\ead{peter.zoltan.csurcsia@vub.be},
\author[Tue,Vub]{Johan Schoukens}\ead{Johan.Schoukens@vub.be},
\author[Tue]{Paul M.J. Van den Hof}\ead{p.m.j.vandenhof@tue.nl}  

\address[Tue]{Department of Electrical Engineering, Eindhoven University of Technology, Eindhoven, The Netherlands}
\address[Vub]{Department of Engineering Technology, Vrije Universiteit Brussel, Belgium}

\begin{keyword}                           
System identification; Interconnected systems; Frequency domain; Dynamic networks; Non-parametric models. 
\end{keyword}                             

\begin{abstract}
In classical approaches of dynamic network identification, in order to identify a system (module) embedded in a dynamic network, one has to formulate a Multi-input-Single-output (MISO) identification problem that requires identification of a parametric model for all the modules constituting the MISO setup including (possibly) the noise model, and determine their model order. This requirement leads to model order selection steps for modules that are of no interest to the experimenter which increases the computational complexity for large-sized networks. 
Also, identification using a parametric noise model (like BJ method) can suffer from local minima, however neglecting the noise model has its impact on the variance of the estimates. 
In this paper, we provide a two-step identification approach to avoid these problems. The first step involves performing a non-parametric indirect approach for a MISO identification problem to get the non-parametric frequency response function (FRF) estimates and its variance as a function of frequency. In the second step, the estimated non-parametric FRF of the target module is smoothed using a parametric frequency domain estimator with the estimated variance from the previous step as the non-parametric noise model. The developed approach is practical with weak assumptions on noise, uses the available toolbox, requires a parametric model only for the target module of interest, and uses a non-parametric noise model to reduce the variance of the estimates. Numerical simulations illustrate the potentials of the introduced method in comparison with the classical identification methods. 
\end{abstract}

\end{frontmatter}

\section{Introduction}
Real-life systems are becoming increasingly complex and largely interconnected, leading to high interest in the field of identification of large-scale interconnected systems known as dynamic networks. These networks can be considered as a set of measurable signals (the node signals) interconnected through linear dynamic systems and can be possibly driven by external excitation signals and/or process noise. The node signals can possibly be measured with sensor/measurement noise. Data-driven modeling in dynamic networks can be broadly classified into full network identification and local module identification. In the former, we deal with identification of the full network dynamics \cite{Haber&Verhaegen:TAC:14,Torres14,Weerts&etal_Autom:18_reducedrank,Weerts&etal_CDC:16,Zorzi&Chiuso:17}, including the aspects of identifiability \cite{Goncalves&Warnick:08, Weerts&etal_Autom:18_identifiability, Hendrickx&Gevers&Bazanella_TAC:19,Bazanella&etal_CDC:17,vanWaarde&Tesi&Camlibel_NECSYS:18,Cheng&etal_CDC:19}, while the latter deals with identifying a particular module (system) in a dynamic network assuming that the topology of the network is given \cite{Chiuso&Pillonetto_Autom:12,VandenHof&etal_Autom:13,Dankers&etal_Autom:15,materassi2015identification,Dankers&etal_TAC:16,Linder&Enqvist_ijc:17,Ramaswamy&etal_CDC:18, Everitt&Bottegal&Hjalmarsson_Autom:18,Gevers&etal:sysid18,VandenHof&etal_CDC:19,Ramaswamy&etal_CDC:19,Ramaswamy&etal_TAC:21,Ramaswamy&etal_Autom:2021,Yu&Verhaegen_TAC:2018}. 

In this paper, we deal with local module identification. In \cite{VandenHof&etal_Autom:13, Dankers&etal_TAC:16}, the classical \emph{direct method} \cite{Ljung:99} for closed loop identification has been generalized to the framework of dynamic network. A \emph{local direct method} has been introduced in \cite{Ramaswamy&etal_TAC:21} for identification of dynamic networks with correlated noise. Indirect identification methods for dynamic networks have been presented in \cite{Gevers&etal:sysid18, Dankers&etal_TAC:16}. An errors-in-variables (EIV) identification framework to the case of dynamic network has been introduced in \cite{Dankers&etal_Autom:15}. The \emph{direct method} \cite{VandenHof&etal_Autom:13} and the simultaneous minimization of prediction error approach \cite{Gunes&etal_IFAC:14} have been extended to a Bayesian setting in \cite{Ramaswamy&etal_CDC:18} and \cite{Everitt&Bottegal&Hjalmarsson_Autom:18} respectively, where a regularized kernel-based method has been used to minimize the mean-squared error of the estimated target module. A \emph{generalized method} that relaxes the sensing and actuation schemes in a dynamic network by combining the framework of direct and indirect method has been introduced in \cite{Ramaswamy&etal_CDC:19}. 
All the aforementioned approaches are time-domain methods. Frequency domain methods for identification in dynamic networks are scarcely explored. A fully non-parametric frequency domain approach has been presented in \cite{Dankers&etal_CDC:15}.

In this paper, we aim at identifying a single module in a dynamic network with reduced complexity. In direct approaches \cite{VandenHof&etal_Autom:13, Dankers&etal_TAC:16, Ramaswamy&etal_TAC:21}, we typically need to perform a MISO or MIMO identification that requires a parameterized model for all modules in  the MISO or MIMO structure. This demands performing a model order selection step using complexity criteria like AIC, BIC, CV also for the modules that are of no interest to the experimenter. For a large-sized network, this task can be computationally demanding (see \cite{Ramaswamy&etal_CDC:18} for an example). Moreover, direct approaches lead to biased estimates in an EIV setting (i.e. dynamic networks with node signals measured with sensor noise) and requires handling of confounding variables\footnote{A confounding variable is an unmeasured variable that induces correlation between the input and output signal of an estimation problem \cite{Pearl:2000}. See \cite{VandenHof&etal_CDC:19} for a formal definition.}. 

Indirect approaches like the \emph{two-stage method} \cite{VandenHof&etal_Autom:13} and the \emph{Instrumental Variable} (IV) method \cite{Dankers&etal_Autom:15} do not require handling of confounding variables and can be used in an EIV setting, but they also face the problem of large number of parameterized models in the MISO structure. It is possible to reduce the complexity a bit by disregarding the parametric noise model in the \emph{two-stage method} \cite{VandenHof&etal_Autom:13} and the IV method \cite{Dankers&etal_Autom:15}, however it comes with the cost of increasing the variance of the estimates. This issue has been dealt for general identification of SISO systems in frequency domain by a semi-parametric approach with a parametric model for the plant and a non-parametric noise model \cite{Schoukens&etal_Autom:97,Schoukens&etal_CDCECC:11}, and has been extended to MIMO systems in \cite{Pintelon&etal_MSSP:2010a,Pintelon&etal_MSSP:2010b}. 

In this paper, we aim at addressing the above problems by developing a semi-parametric frequency domain approach that requires a parametric model only for the target module of interest and incorporates noise modeling through a non-parametric noise model. We achieve this using a two-step approach, where the first step involves a MISO identification problem to get the non-parametric frequency response function (FRF) estimate of the target module and also its variance, thereby avoiding parametric models in the MISO setup. In the second step, the FRF estimate of the target module is smoothed using a parametric frequency domain estimator with the estimated variance from the previous step as the noise model, thereby reducing the variance of the estimate. The approach is practical, implemented using the already available MATLAB toolbox and can be implemented for a network with correlated noise and an EIV setting. Numerical simulations are performed for a dynamic network example to show the effectiveness of the developed method compared with the classical methods for dynamic network identification.

\section{Problem statement}\label{sec:prob stat}
Following the setup of \cite{VandenHof&etal_Autom:13}, we consider a dynamic network that is built up of $L$ scalar measurable internal variables or nodes $w_j(t)$, $j$ = $1, \dots, L$. Each internal variable is described as:
\begin{align}
w_j(t) = \sum_{l \in \mN_j}
G_{jl}^0(q)w_l(t) + r_j(t) + v_j(t)
\label{eq:netw_def}
\end{align}
where,
\begin{itemize}
	\item $q^{-1}$ is the shift (delay) operator i.e. $q^{-1}w_j(t) = w_j(t-1)$;
	\item $\mathcal{N}_j$ is the set of node indices $k$ such that $G_{jl} \nequiv 0$;
	\item $G_{jl}^0(q)$ are proper rational transfer function for $j = 1, \dots, L$ and $k = 1, \dots, L$;
	\item There are no self-loops in the network, i.e. nodes are not directly connected to themselves $G_{jj} = 0$;
	\item $v_j(t)$ is the \emph{process noise}, where the vector process $v=[v_1 \cdots v_L]^T$ is modelled as a stationary stochastic process with rational spectral density $\Phi_v(\omega)$, such that there exists a white noise process $e:= [e_1 \cdots e_L]^T$, with covariance matrix $\Lambda>0$ such that
	$ v(t) = H(q)e(t)$,
	where $H(q)$ is monic, square, stable and minimum-phase. The situation of correlated noise refers to the situation that $\Phi_v(\omega)$ and $H$ are non-diagonal;
	\item $r_j(t)$ is a measured external excitation signal entering node $w_j(t)$. In some nodes, it may be absent.
\end{itemize}
On combining the equation for all nodes in the network, we obtain the network equation (time and frequency dependence is omitted below),
\begin{equation}\label{eq:networkmodel}
\begin{split}
\begin{bmatrix}
w_1 \\ w_2\\ \vdots \\ w_L
\end{bmatrix} &\!\!=\!\!
\begin{bmatrix}
\!\!0 & \!\!G_{12}^0(q)& \!\!\dots & \!\!G_{1L}^0(q)\\
\!\!G_{21}^0(q) & \!\!0& \!\!\dots & \!\!G_{2L}^0(q)\\
\!\!\vdots & \!\!\ddots & \!\!\ddots &\!\!\vdots \\
\!\!G_{L1}^0(q) & \!\!G_{L2}^0(q)& \!\!\dots & \!\!0\\
\end{bmatrix}\!\!\!\begin{bmatrix}
w_1 \\ w_2\\ \vdots \\ w_L
\end{bmatrix} \!\!+\!\!
\begin{bmatrix}
r_1 \\ r_2\\ \vdots \\ r_L
\end{bmatrix} \!\!+\!\!
\begin{bmatrix}
v_1 \\ v_2\\ \vdots \\ v_L
\end{bmatrix}\\
&= G^0(q)w(t) + r(t) + v(t)\\
\end{split}
\end{equation}
The representation in \eqref{eq:networkmodel} is an extension of the Dynamic Structure Function (DSF) representation \cite{Goncalves&Warnick:08}. The dynamic network is assumed to be stable, i.e. $(I - G^0(q))^{-1}$ is stable, and well posed (see \cite{VandenHof&etal_Autom:13} for details). 

Every node signal is measured using a sensor, due to which there can be measurement errors in the recorded measured node signal of $w_k$. Therefore the measurement of $w_k$ accounting for the sensor noise is given by $\tilde w_k = w_k + s_k$, where $s_k$ is the sensor noise and is modelled as a stochastic process with rational power spectral density. This will be the EIV setting in dynamic networks. The identification problem that we consider in this paper is to identify the target module of interest $G_{ji}^0$ from measurements of $w$ and possibly $r$. To this end, we choose a parameterization of $G_{ji}^0(q)$, denoted as $G_{ji}(q,\theta)$, that describes the dynamics of the module of interest for a certain parameter vector $\hat \theta = \theta_0 \in \mathbb{R}^{n_\theta}$.

\section{The direct approach}\label{sec:Direct_Id}
\subsection{The standard direct method}
The equation \eqref{eq:netw_def} represents a MISO structure and is the starting point of the direct method in \cite{VandenHof&etal_Autom:13}. Let $\mathcal{D}_j$ denote the set of indices of the internal variables that
are chosen as predictor inputs and let $\mR$ denote the non-zero external signals in $r$. Let $w_{\smD_j}$ denote the vector $[\begin{matrix}
w_{k_1} & \dots & w_{k_n}
\end{matrix}]^\top$, where $\{k_1,\dots, k_n\} = \mD_j$ and let $r_{\smR}$ denote the vector $[\begin{matrix}
r_{k_1} & \dots & r_{k_n}
\end{matrix}]^\top$, where $\{k_1,\dots, k_n\} = \mR$. In the standard direct method for dynamic networks \cite{VandenHof&etal_Autom:13}, identification of a module $G_{ji}$ can be done by selecting a set of predictor inputs $\mathcal{D}_j$ such that $i \in \mathcal{D}_j$, and subsequently estimating a multiple-input single output model for the transfer functions in $G_{j\smallsub{D}_j}$. This can be done by considering the one-step-ahead predictor\footnote{$\Eb$ refers to $\lim_{N\rightarrow\infty} \frac{1}{N} \sum_{t=1}^N \E$, and $w_j^{\ell}$ and $w_{\mathcal{D}_j}^{\ell}$ refer to signal samples $w_j(\tau)$ and $w_k(\tau)$, $k\in \mathcal{D}_j$, respectively, for all $\tau \leq \ell$.} $\hat w_j(t|t-1):= \Eb \{w_j(t)\ |\ w_j^{t-1},w_{\mathcal{D}_j}^t\}$, and the resulting prediction error (\cite{Ljung:99}) $\varepsilon_j(t,\theta) =
w_j(t) - \hat{w}_j(t|t-1,\theta)$, given by
\vspace{-0.15cm}
\begin{align} \label{eq.predictionError}
\varepsilon_j(t,\theta) =  H_j(\theta)^{-1} \Big ( w_j - \!\! \sum_{k \in \mathcal{D}_j}
G_{jk}(q,\theta) w_k -  r_j \Big)
\end{align}
\noindent where arguments $q$ and $t$ have been dropped for notational clarity. The parameterized transfer functions $G_{jk}(q,\theta)$, $k \in
\mathcal{D}_j$ and
$H_j(\theta)$ are estimated by
minimizing the sum of squared (prediction) errors:
$V_j(\theta) = \frac{1}{N}
\sum_{t=0}^{N-1} \varepsilon_j^2(t,\theta),
$
where $N$ is the length of the data set. Let $\hat{\theta}_N$ denote the minimizing argument of $V_j(\theta)$. We note that in the above formulation, the prediction error depends also on the additional parameters entering the modules that are not of interest, and on the noise model, which needs to be identified to guarantee consistent estimates of $\theta$. Therefore, all these modules require a model order selection step that could have a detrimental effect in large-scale networks due to computational complexity. 

\subsection{Predictor input selection}
Predictor input selection (i.e. selecting $\mD_j$) plays an important role in the direct method, for guaranteeing that the identified transfer function is an estimate of the target module $G_{ji}^0$ and not a biased estimate of the target module. We call this as \emph{module invariance} in networks. In \cite{VandenHof&etal_Autom:13}, the set $\mD_j$ is chosen to be $\mN_j$ (i.e. all in-neighbors of the output of the target module). However, it is possible to choose a subset of $\mN_j$ in $\mD_j$ as predictor inputs provided certain conditions are satisfied \cite{Dankers&etal_TAC:16}. The main condition is given in Property \ref{propt1}.
\begin{property}\label{propt1}
	To identify a target module $G_{ji}$, consider a set of internal variables $w_k, k \in \mD_j$. Let $\mD_j$ satisfy the following properties:
	\begin{enumerate}
		\item $i \in \mD_j$ and $j \notin \mD_j$;
		\item Every path from $w_i$ to $w_j$, excluding the path through $G_{ji}$, pass through a node $w_k, k \in \mD_j$ (\emph{parallel path condition});
		\item Every loop through $w_j$ pass through a node $w_k, k \in \mD_j$ (\emph{loop condition}).
	\end{enumerate}
\end{property}
When the above property is satisfied, the \emph{direct method} provides a consistent estimate of the target module, if data informativity conditions are satisfied, and in addition there are no confounding variables for the estimation problem $w_{\smD_j} \rightarrow w_{j}$. We use the identification setup of the direct method as the basis of the approach developed in this paper.

\section{The frequency domain approach}
\subsection{Introduction}
A special semi-parametric two-step indirect approach can also be used for the identification of a single local module. In the first step, a nonparametric indirect method is used to get FRF and variance estimates of the local module that is involved in the problem. In a second step, the estimated nonparametric FRF of the local module of interest is smoothed using a parametric frequency domain estimator using the estimated variance as a nonparametric noise model. \par
The semi-parametric approach offers two major advantages. First, the method complexity is independent of the complexity of the network. The complexity of the first non-parametric step is set by the number of local modules that need to be estimated in order to isolate the local module of interest. Second, in the parametric step, a parametric model is only estimated for the local module of interest, in other words, the other modules are not entering into the problem. Moreover, only the plant model needs to be identified because a nonparametric noise model that is obtained in the first step is used as a frequency weighting.

\subsection{Nonparametric FRF estimation}
In this section we will use a special implementation of the Local Polynomial Method (LPM)  \cite{Pintelon&etal_MSSP:2010a} for estimating the FRF of the target module. The basic idea of LPM is to cope with leakage and transient issues and it brings two advantages:
\begin{itemize}
	\item The approach is not restricted to periodic excitations, it works well with noise excitation as well. Of course, the advantages of using a periodic excitation still hold true and we still advocate to use periodic excitations whenever this is possible.
	\item The LPM estimate provides an FRF estimate for the modules together with their co-variance matrices (as a function of the frequency). In this step there is almost no user interaction needed, and the complexity of the identification problem is (almost) independent of the size of the network.
\end{itemize}
We consider a straightforward, iteration free, direct implementation of LPM calculated per excited frequency lines as discussed in \cite{CSURCSIA2020106926} and implemented in the SAMI toolbox (Simplified Analysis for Multiple Input \cite{Csurcsia_2020}). This toolbox offers a user friendly environment (GUI or Command line driven) to make a full nonparametric analysis of MIMO (non)linear systems. \par
 Next, the LPM method is briefly elaborated. The LPM relies on the following frequency domain nonparametric baseline model at the excited frequency lines, for a generic input-output relationship:
\begin{align}
Y(k) = G(k)U(k)+T(k)
\label{eq:lpm_baseline}
\end{align}
where,
\begin{itemize}
	\item $k$ represents an excited frequency line (bin)
	\item $Y$ is the output signal in the frequency domain with $n_y$ outputs (dimension of $Y$ is $n_y \times N$, dimension of $Y(k)$ is $n_y \times 1$);
	\item $U$ is the excitation signal in the frequency domain with $n_i$ inputs (dimension of $U$ is $n_{i} \times N$, dimension of $U(k)$ is $n_i \times 1$);
	\item $T$ represents the (autonomous) transient term (dimension of $T$ is $n_y \times N$, dimension of $T(k)$ is $n_y \times 1$)
\end{itemize}
Because the FRF and the transient are smooth, it is possible to use polynomial approximations such as the LPM.  In the process, a (narrow) sliding processing window is used with a polynomial degree of $d$. First, an excited frequency line is selected (denoted by frequency index $f$), this is called the central frequency. This is the middle point of the processing window. Around this frequency line, in a $\pm d/2$ radius, a narrow band is selected such that the $r$ (radius) is given by 
\begin{align}
r=\floor{-\frac{d}{2}}\dots 0\dots \ceil{\frac{d}{2}}.
\label{eq:lpm_r}
\end{align}
In this band ($f+r$) all the (excited and non-excited) frequency lines are used to estimate the transfer functions in polynomial form. The measured output is at the excited frequency index $f$ in a radius of $r$ as follows:
\begin{align}
Y(f+r)=G_p(f+r)U(f+r)+T_p(f+r)+E(f+r)
\label{eq:lpm_output}
\end{align}
where $G_p$ and $T_p$ are polynomials of order $d$. The polynomials are given by:
\begin{align}
G_p=\sum_{s=0}^{d}g_s r^s=g_0+g_1 r+\dots+g_d r^d 
\label{eq:lpm_gp}
\end{align}
\begin{align}
T_p=\sum_{s=0}^{d}t_s r^s=t_0+t_1 r+\dots+t_d r^d 
\label{eq:lpm_gt}
\end{align}
The LPM parameters at the center frequency $f$ can be estimated by the following LS cost function:
\begin{align}
\hat{\eta}(f)=\underset{\eta(f)}{\mathrm{argmin}} \sum_{r} |Y(f+r)-G_p(f+r)U(f+r)\nonumber\\
-T_p(f+r) |^2 
\label{eq:lpm_cf}
\end{align}
where $\hat{\eta}(f)$ represents the estimates of the LPM polynomials (i.e. the terms in (\ref{eq:lpm_gp},\ref{eq:lpm_gt})) at frequency index $f$. In (\ref{eq:lpm_gp}) the estimated value of $g_0$ corresponds to $G$, i.e. the quantity of interest. \par
The LS problem stated in (\ref{eq:lpm_cf}) is recommended to be solved output channel per output channel. In this case, for each frequency line there are $(d+1)(n_i+1)$ unknown parameters to be estimated; ($(d+1)n_i$ in $G_p$, and $d+1$ in $T_p$). This means, for the algorithm to work, the bandwidth $r$ should be at least as wide as the number of unknown parameters. In this work we consider second degree polynomials ($d=2$). The restriction on the minimum bandwidth also implies that in the FRF measurements, in the (-3 dB) bandwidth of the lowest damped mode (the sharpest peak in the FRFs) should be at least $(d+1)(n_i+1)$ points measured. By further increasing the bandwidth of LPM, an increased smoothing of the nonparametric estimate over the frequency is obtained. However, a too large bandwidth will create a bias. 
Since the FRFs will be further smoothed in the second parametric step, the perfectly chosen parameters of the smoothing processing is less crucial.  
Note that (\ref{eq:lpm_cf}) must be solved with the help of a numerically stable inversion method. 

\subsection{The nonparametric indirect method}
The previous subsection contains the formal definition of the LPM method. In the dynamic network setup, an indirect method is used. \par
First, the transfer is estimated between the non-zero external excitation signals $r_k, k \in \mR$ and the nodes $w_k, k \in \mD_j$. This is reflected in the frequency domain through the following MIMO problem:
\begin{align}
W_{\smD_j}(f+r) = S_{\smD_j\smR}(f+r)R_{\smR}(f+r)+T(f+r)
\label{eq:lpm_indirect_step1}
\end{align}
where,
\begin{itemize}
	\item $W_{\smD_j}$ is the node value in the frequency domain for the node signals in $w_{\smD_j}$;
	\item $R_{\smR}$ represents the reference signals $r_{\smR}$ in the frequency domain;
	\item $S_{\smD_j\smR}$ is the Frequency Response Matrix (FRM) i.e. collection of FRFs, between the reference signals in $r_{\smR}$ and node signals $w_{\smD_j}$;
	\item $T$ represent the (autonomous) transient term.
\end{itemize}
This problem is solved with the help the proposed LPM method such that the estimated FRM $\hat{S}_{\smD_j\smR}$ is obtained between the reference signals $r_{\smR}$ and the nodes $w_{\smD_j}$. Note, that for a unique solution, it is required that $R_{\smR}$ is full rank.\par

The next step is needed to reduce the cross-correlation problems. In place of the measured nodes $W_{\smD_j}$ we will use the simulated node values $\hat{W}_{\smD_j}$ using $\hat{S}_{\smD_j\smR}$:
\begin{align}
\hat{W}_{\smD_j}(f+r) = \hat{S}_{\smD_j\smR}(f+r)R_{\smR}(f+r).
\label{eq:lpm_indirect_step3}
\end{align}
The following step is to estimate the transfer between $\hat{W}_{\smD_j}$ and the output of the target module using the following relationship:
\begin{align}
W_{j}(f+r) = G_{j\smD_j}(f+r)\hat{W}_{\smD_j}(f+r) + T(f+r)
\label{eq:lpm_indirect_step4}
\end{align}
where
\begin{itemize}
	\item $W_{j}$ is the measured node value of the output of the target module in the frequency domain;
	\item $G_{j\smD_j}$ is the FRM between the target module and the nodes in $w_{\smD_j}$.
\end{itemize}
This problem should be solved again with the help of the proposed LPM method such that the estimated FRM $\hat{G}_{j\smD_j}$ is obtained. We can now isolate the local module $\hat{G}_{ji}$ in the FRM estimate $\hat{G}_{j\smD_j}$.

\subsection{Parametric processing}
One must post-process the non-parametric local module $\hat{G}_{ji}$ in order to obtain the parametric rational transfer function form $\hat{G}_{ji}(\theta)$. In this work, a weighted least squares estimation is used for the identification:
\begin{align}
V(\theta)=\frac{1}{F}\sum_{f}\frac{|\hat{G}_{ji}(f)-\hat{G}_{ji}(f|\theta)|^2}{{\hat{\sigma}^2}_{\hat{G}_{ji}}(f)}
\label{eq:cf_elvis}
\end{align}
where
\begin{itemize}
	\item $\theta$ represents the parameters in the parametric local module estimate;	
	\item $F$ is the number of frequency components;
	\item ${\hat{\sigma}^2}_{\hat{G}_{ji}}$ is the non-parametric co-variance estimate of the the local module estimate obtained from the LPM method \cite{CSURCSIA2020106926, Pintelon&etal_MSSP:2010a}.
\end{itemize}
In order to minimize the cost function given in (\ref{eq:cf_elvis}) the FDIDENT toolbox is used \cite{fdident}.

\subsection{Reflection}
The presented method combines a two-stage indirect method for estimating a local module, and combines this with a non-parametric noise model to reduce the variance of the final parametric module estimate. The use of the non-parametric noise model is particularly enabled by the fact that the problem is addressed in the frequency domain. Additionally there is no need for parametrically modeling nuisance modules, typically present in MISO network estimation problems, and therefore extensive model order selection procedures are avoided.

\section{Numerical Simulations}\label{sec:Num_sim}
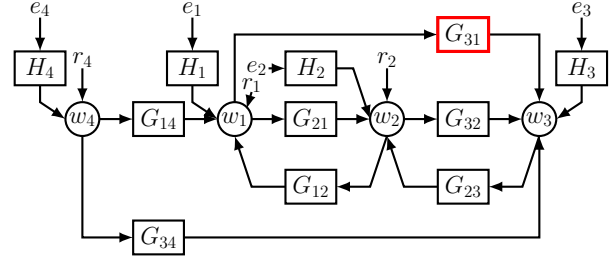
\begin{figure}
	\centering
	\usetikzlibrary{arrows}
	\begin{tikzpicture}[thick,scale=0.45, every node/.style={scale=0.45}]
	\draw  (-4.5,4.5) rectangle (-3,3.5) node[pos = 0.5] {\LARGE $G_{21}$};
	\draw [-latex](-5.5,4) -- (-4.5,4);
	\draw [-latex] (-6,4) ellipse (0.5 and 0.5) node{\LARGE $w_1$};
	\draw [-latex](-7.5,4) -- (-6.5,4);
	\draw [-latex] (-9,4.5) rectangle (-7.5,3.5) node[pos = 0.5] {\LARGE $G_{14}$};
	\draw [-latex](-10,4) -- (-9,4);
	\draw [-latex] (-10.5,4) ellipse (0.5 and 0.5) node{\LARGE $w_4$};
	\draw [-latex](-3,4) -- (-2,4);
	\draw [-latex] (-1.5,4) ellipse (0.5 and 0.5) node{\LARGE $w_2$};
	\draw [-latex](-1,4) -- (0,4);
	\draw [-latex] (0,4.5) rectangle (1.5,3.5) node[pos = 0.5] {\LARGE $G_{32}$};
	\draw [-latex](1.5,4) -- (2.5,4);
	\draw [-latex] (3,4) ellipse (0.5 and 0.5) node{\LARGE $w_3$};
	\draw [-latex](-10.5,5.5) node at (-10.5,5.75) {\LARGE $r_4$}-- (-10.5,4.5);
	\draw [-latex](-1.5,5.5) node at (-1.5,5.75) {\LARGE $r_2$}-- (-1.5,4.5);
	\draw [-latex] (-4.5,2.5) rectangle (-3,1.5) node[pos = 0.5] {\LARGE $G_{12}$};
	\draw [-latex] (0,2.5) rectangle (1.5,1.5) node[pos = 0.5] {\LARGE $G_{23}$};
	\draw [-latex](-1.5,3.5) -- (-2,2) -- (-3,2);
	\draw [-latex](3,3.5) -- (2.5,2) -- (1.5,2);
	\draw [-latex](0,2) -- (-1,2) -- (-1.5,3.5);
	\draw [-latex](-4.5,2) -- (-5.5,2) -- (-6,3.5);
	\draw [-latex] (-9,1) rectangle (-7.5,0) node[pos = 0.5] {\LARGE $G_{34}$};
	\draw [-latex](-10.5,3.5) -- (-10.5,0.5) -- (-9,0.5);
	\draw [-latex](-7.5,0.5) -- (3,0.5) -- (3,3.5);
	\draw [-latex] (-12.5,6) rectangle (-11,5) node[pos = 0.5] {\LARGE $H_{4}$};
	\draw [-latex] (0,7) rectangle (1.5,6) node[pos = 0.5] {\LARGE $G_{31}$};
	\draw [-latex](-6,4.5) -- (-6,6.5) -- (0,6.5);
	\draw [-latex](1.5,6.5) -- (3,6.5) -- (3,4.5);
	\draw [-latex] (-4.5,6) rectangle (-3,5) node[pos = 0.5] {\LARGE $H_{2}$};
	\draw [-latex] (-6.5,6) rectangle (-8,5) node[pos = 0.5] {\LARGE $H_{1}$};
	\draw [-latex] (3.5,6) rectangle (5,5) node[pos = 0.5] {\LARGE $H_{3}$};
	\draw [-latex](-11.75,5) -- (-11.75,4.5) -- (-11,4);
	\draw [-latex](-7.25,5) -- (-7.25,4.5) -- (-6.5,4);
	\draw [-latex](-3,5.5) -- (-2.5,5.5) -- (-2,4);
	\draw [-latex](-5,5.5) node at (-5.35,5.5) {\LARGE $e_2$} -- (-4.5,5.5);
	\draw [-latex](-11.75,7) node at (-11.75,7.25) {\LARGE $e_4$}-- (-11.75,6);
	\draw [-latex](-7.25,7) node at (-7.25,7.25) {\LARGE $e_1$} -- (-7.25,6);
	\draw [-latex](4.25,5) -- (4.25,4.5) -- (3.5,4);
	\draw [-latex](4.25,7) node at (4.25,7.25) {\LARGE $e_3$} -- (4.25,6);
	\draw [-latex, red, very thick] (0,7) rectangle (1.5,6);
	\draw [-latex](-5.5,4.8) node at (-5.5,5) {\LARGE $r_1$}-- (-5.65,4.3);
	\end{tikzpicture}
	\caption{Network example with 4 internal nodes, 3 reference signals and a noise sources at each node.}
	\label{fig:dynnet_Ex_wnoise1}
\end{figure}

Numerical simulations are performed with a dynamic network example as given in Figure \ref{fig:dynnet_Ex_wnoise1}, to evaluate the performance of the developed method discussed in Section 4, which we abbreviate as ``iLPM + ELiS" (indirect Local Polynomial Method combined with ELiS cost function). We compare the method with the standard direct method ``DM+TO" \cite{VandenHof&etal_Autom:13} and the Empirical Bayes Direct Method (``EBDM") \cite{Ramaswamy&etal_CDC:18}. The goal is to identify $G_{31}^0$. The network example is the same network as in \cite{Ramaswamy&etal_Autom:2021}, with the only difference in the presence of $r_1$ signal. The network modules in Figure \ref{fig:dynnet_Ex_wnoise1} are given by,
\begin{align*}
&G_{31}^0 = \frac{q^{-1} + 0.05q^{-2}}{1 + q^{-1} + 0.6q^{-2}} = \frac{b_1^0q^{-1} + b_2^0q^{-2}}{1 + a_1^0q^{-1} + a_2^0q^{-2}}\\
&G_{32}^0 = \frac{0.09 q^{-1}}{1 + 0.5 q^{-1}};\\
&G_{34}^0 = \frac{1.184 q^{-1} - 0.647 q^{-2} + 0.151 q^{-3} - 0.082 q^{-4}}{1 - 0.8 q^{-1} + 0.279 q^{-2} - 0.048 q^{-3} + 0.01 q^{-4}};\\
&G_{14}^0 = G_{21}^0 = \frac{0.4q^{-1} - 0.5q^{-2}}{1 + 0.3q^{-1}};H_{1}^0 = \frac{1}{1 + 0.2q^{-1}};\\
&G_{12}^0 = G_{23}^0 = \frac{0.4q^{-1} + 0.5q^{-2}}{1 + 0.3q^{-1}};H_{2}^0 = \frac{1}{1 + 0.3q^{-1}}\\
&H_{3}^0 = \frac{1 - 0.505 q^{-1} + 0.155 q^{-2} - 0.01 q^{-3}}{1 - 0.729 q^{-1} + 0.236 q^{-2} - 0.019 q^{-3} }; H_{4}^0 = 1.
\end{align*}

We run $100$ independent Monte Carlo experiments where the data is generated using known reference signals $r_1(t)$, $r_2(t)$ and $r_4(t)$ that are realizations of independent white noise with variance of 0.1. The number of data samples is $N$ = 500. The noise sources $e_1(t)$, $e_2(t)$, $e_3(t)$ and $e_4(t)$ have variance 0.05, 0.08, 1, 0.1, respectively. We assume that we know the model order of $G_{31}^0(q)$. In the case of the direct method, we solve a 3-input/1-output MISO identification problem with $w_1(t)$, $w_2(t)$ and $w_4(t)$ as inputs. 
For the direct method, we consider that the model orders of all modules in the MISO setup are known. The EBDM \cite{Ramaswamy&etal_CDC:18} considers the target module $G_{31}^0$ in the above MISO identification problem as parameterized and the rest of the modules (nuisance modules) in the MISO setup are modeled as zero mean independent Gaussian processes, whose covariance matrices are described by a \emph{stable spline kernel}  which encodes stability and smoothness of its impulse response. Each nuisance module is described only by two hyperparameters of the kernel, thereby reducing the number of parameters. The model order of the target module is considered to be known. The length of the impulse response of each module in the EBDM is considered to be $\ell = 50$. The indirect LPM (iLPM) method uses a second order polynomial. For this configuration the minimal required bandwidth should be at least 12 (i.e. $(d+1)(n_i+1)=3\times4$). Due to the excessive noise on the signals, the bandwidth should be higher. Using a simple Least Squares fitting method, the bandwidth has been set to 24, which is sufficient to reduce the effect of the noise and still give a small bias error.  

\begin{figure}
	\centering
	\includegraphics[scale=0.22]{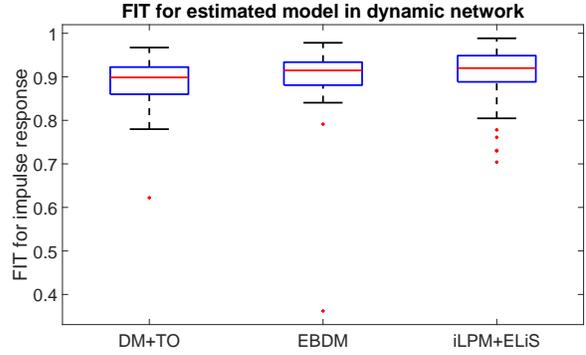}
	\caption{Box plot of the fit of the impulse response of $\hat{G}_{31}$ obtained by the developed method, Direct method and EBDM. Number of data samples used for estimation is $N$ = 500.}
	\label{fig:boxplotimp}
\end{figure}

\begin{figure}
	\centering
	\includegraphics[scale=0.22]{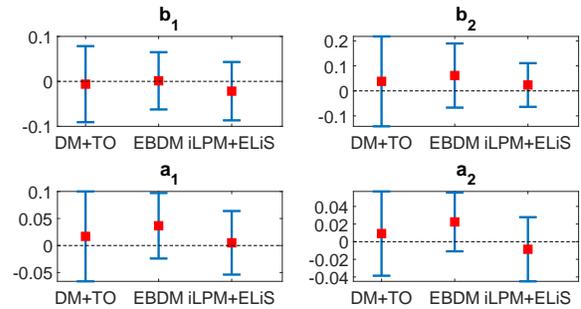}
	\caption{Bias and standard deviation of each parameter obtained from 100 MC simulations using different identification methods.}
	\label{fig:errorplotmse}
\end{figure}

\begin{figure}
	\centering
	\includegraphics[scale=0.22]{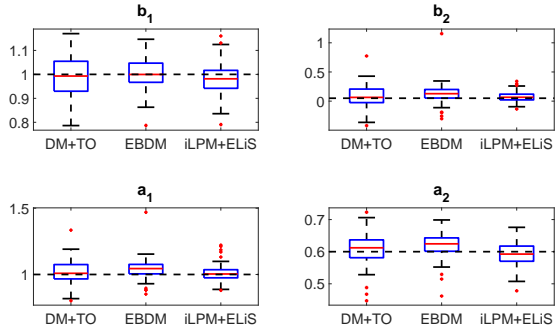}
	\caption{Box plot of the parameters of $\hat{G}_{31}$ obtained by the developed method, Direct method and EBDM. Number of data samples used for estimation is $N$ = 500.}
	\label{fig:boxplotpara}
\end{figure}

\begin{figure}
	\centering
	\includegraphics[scale=0.22]{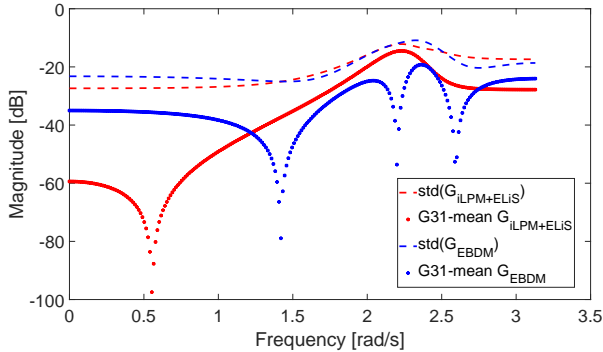}
	\caption{FRF plot showing 1) the error between the true value and the mean value (dotted lines) and 2) the standard deviation on the mean of the FRF (dashed lines), for the developed method (red) and EBDM (blue) calculated from 100 realizations.}
	\label{fig:FRFplot}
\end{figure}

To evaluate the performance of the methods, we use the standard goodness-of-fit metric,
\beq
\textrm{Fit} = 1 - \frac{{\norm{g_{ji}^0 - \hat g_{ji}}}_2}{{\norm{g_{ji}^0 - \bar g_{ji}}}_2}, \nonumber
\eeq
where $g_{ji}^0$ is the true value of the impulse response of $G_{ji}^0$, $\hat g_{ji}$ is the impulse response of the estimated target module and $\bar g_{ji}$ is the vector with the sample mean of values in $g_{ji}^0$ as its elements. The box plots of the fits of the estimated impulse response of $G_{31}(q)$ are shown in Figure \ref{fig:boxplotimp}, where we have compared the Direct method with true model orders (`DM+TO'), the Empirical Bayes Direct Method (`EBDM'), and the developed method (`iLPM+ELiS'). It can be noted that the median of the fit of the developed method is higher that the other considered methods.
Figure \ref{fig:errorplotmse} shows the mean and standard deviation of the parameter estimates of $G_{31}$.
It is evident that the developed method (`iLPM+ELiS') gives a smaller bias and a reduced variance (in particular when considering parameter parameter $b_2$) compared to the other considered identification methods. This can also be confirmed from the box plot of the parameters of $\hat{G}_{31}$ in figure \ref{fig:boxplotpara}.
The reduction in variance is attributed to the implementation of a non-parametric noise model and it can be witnessed that it is on par with the EBDM which is expected to provide reduced variance due to incorporation of regularization in its approach. 
The mean and standard deviation on the mean of the FRF estimated from 100 MC simulations is shown in figure \ref{fig:FRFplot}. It can be observed frequency response plot that the error between the true value and the mean value is far below the standard deviation, indicating that the developed method is unbiased. 

Considering a relatively small sized network with 3 modules in the MISO structure, the developed method proves effective. When the size of the network grows, the results of the direct method may deteriorate further due to increase in variance; furthermore, it is expected that in large networks the model order selection step contributes to inaccurate results. The developed method requires a parametric model only for the target module and circumvents the problem of model order selection. Also, the incorporation of a non-parametric noise model substantially reduces the variance in the target module estimate and also avoids the problem of local minima that can occur in a parametric noise model as in the direct method. Thus the developed method is expected to serve as an effective local module identification method also in large dynamic networks.

\begin{remark}
 Being an indirect approach, the developed method also provides consistent estimates in an EIV setting \cite{VandenHof&etal_Autom:13}, thereby allowing our measurements to be distorted by sensor noise. 
\end{remark}

\section{Conclusions}
A two-step indirect local module identification method has been developed in this paper. The developed method (`iLPM+ELiS') avoids the problem of model order selection for all the modules that are of not interest to the experimenter, but still needs to be estimated in order to obtain consistent estimate of the target module. Due to the incorporation of a non-parameteric noise model, we achieve a clear benefit in the mean square error of the estimated target module and also the benefit of avoiding possible local minima when a parameteric noise model in used. Therefore, the developed method is less complex and scalable to large sized networks. Furthermore, the method is built on using the already available toolbox SAMI and FDIDENT, making it practical and user-oriented. Numerical simulations performed with a dynamic network example illustrate the potentials of the developed method on comparison with already available methods. The developed method provide target module estimates with greatly reduced bias and variance on comparison with the other compared methods for local module identification.

\bibliography{Paul_Dynamic_Networks_Library}
\bibliographystyle{plain}

\end{document}